\documentclass[amsmath,amssymb]{revtex4}
\begin{document}
\hspace*{4.5 in}CUQM-124

\vspace*{0.3 in}
\title{The Klein-Gordon equation with the Kratzer potential in $d$ dimensions}
\author{Nasser Saad}
\email{nsaad@upei.ca}
\affiliation{Department of Mathematics and Statistics,
University of Prince Edward Island,
550 University Avenue, Charlottetown,
PEI, Canada C1A 4P3.}
\author{Richard L. Hall}%
 \email{rhall@mathstat.concordia.ca}
\affiliation{Department of Mathematics and Statistics, Concordia University,
1455 de Maisonneuve Boulevard West, Montr\'eal,
Qu\'ebec, Canada H3G 1M8}%
\author{Hakan Ciftci}
 \email{hciftci@gazi.edu.tr}
\affiliation{
Gazi Universitesi, Fen-Edebiyat Fak\"ultesi, Fizik
B\"ol\"um\"u, 06500 Teknikokullar, Ankara, Turkey.
}%
\def\dbox#1{\hbox{\vrule  
                        \vbox{\hrule \vskip #1
                             \hbox{\hskip #1
                                 \vbox{\hsize=#1}%
                              \hskip #1}%
                         \vskip #1 \hrule}%
                      \vrule}}
\def\qed{\hfill \dbox{0.05true in}}  
\def\square{\dbox{0.02true in}} 
\begin{abstract}
\noindent{\bf Abstract:} We apply the Asymptotic Iteration Method to obtain the bound-state energy spectrum for the $d$-dimensional Klein-Gordon equation with scalar $S(r)$ and vector potentials $V(r)$. When $S(r)$ and $V(r)$ are both Coulombic, we obtain all the exact solutions; when the potentials are both of Kratzer type, we obtain all the exact solutions for $S(r)=V(r)$; if $S(r) > V(r)$ we obtain exact solutions under certain constraints on the potential parameters: in this case, a possible general solution is found in terms of a monic polynomial, whose coefficients form a set of elementary symmetric polynomials.

\end{abstract}
\vskip0.1 true in
\maketitle
\section{Introduction}	
\noindent The search for exact solutions of wave equations, whether non-relativistic or relativistic, has been an important research area since the birth of quantum mechanics. Recently the Asymptotic Iteration Method (AIM) has received much attention as a method for solving the Schr\"odinger equation~\cite{Hakan}-\cite{barakat3}, both analytically and approximately. It has been applied to a large number of physically interesting potentials and has often yielded highly-accurate results. Very recently, AIM has been used to study the bound states of the Klein-Gordon and Dirac equations for a number of special potentials \cite{ciftci}-\cite{durmus}. In the case of the Klein-Gordon equation, AIM was used to study the bound-states in the case of equal vector and scalar potential. To our knowledge, the method has never been used to study the bound-states of the Klein-Gordon equation in the case of unequal vector and scalar potentials. In the present paper, we have adapted the method to treat problems where the unequal vector and scalar potentials are of Coulomb or Kratzer type. Our main goal is to investigate the exact solutions, whether the scalar and vector potentials, $S(r)$ and $V(r)$, are equal or not. In the case of equal vector and scalar potentials, it is known that the Klein-Gordon equation usually reduces to a Schr\"odinger-type equation, which can be studied, for example, by use of the Nikiforon-Uvarov method \cite{berkdemir}, or by transforming the equation into a classical hypergeometric differential equation with known solution~\cite{qqiang}-\cite{qiang}. It is known~~\cite{greiner}-\cite{dom} that when $S(r)\geq V(r)$, bound-state solutions exist. Usually the case when the scalar potential is equal to the vector potential is considered separately. The advantage of AIM is that it allows a unified approach that can be used to study the bound-state solutions with equal or unequal scalar and vector potentials. 
\vskip 0.1true in
\noindent  The Klein Gordon equation has received considerable attention in the literature \cite{dom}-\cite{Guyou}. The present paper is organized as follows. Section~II is devoted to a brief introduction to the Klein-Gordon equation in $d$-dimensions. In section~III we summarize the Asymptotic Iteration Method (AIM) \cite{Hakan}. In Section~IV we use AIM to find detailed analytic solutions to the Klein-Gordon equation with Coulomb potentials in $d$-dimensions.  The lowest even-parity solution for the Coulomb problem in one dimension is paradoxical and is still under study \cite{case}-\cite{hallkg}. Our solutions in section IV for $d=1$ confirm the exact solutions of the Klein Gordon equation with mixed vector and scalar Coulomb potentials on the half-line obtained earlier by de Castro \cite{decastro}. In Section~V, we use AIM to derive exact solutions with equal and unequal scalar and vector Kratzer-type potentials.  Our conclusions are presented in section VI.

\section{The Klein-Gordon equation in $d$ dimensions} 
\noindent The $d$-dimensional Klein-Gordon equation for a particle of mass $M$ with radially symmetric Lorentz vector and Lorentz scalar potentials,  $V(r)$ and $S(r)$, $r = \|{\mathbf r}\|,$ is given (in atomic units $\hbar=c=1$) \cite{greiner,alhaidari} by
\begin{equation}\label{eq1}
\{-\Delta_d+[M+S(r)]^2\}\Psi({\bf r})=[E-V(r)]^2\Psi({\bf r}),
\end{equation}
where $E$ denotes the energy and $\Delta_d$ is the $d$-dimensional Laplacian. Transforming to the $d$ dimensional spherical coordinates $(r,\theta_1\dots\theta_{D-1})$, the variables can be separated using
\begin{equation}\label{eq2}
\Psi({\bf r})=R(r)Y_{l_{d-1},\dots,l_1}(\theta_1\dots\theta_{d-1})
\end{equation}
where $R(r)$ is a radial function, and $Y_{l_{d-1},\dots,l_1}(\theta_1\dots\theta_{d-1})$ is a normalized  hyper-spherical harmonic with eigenvalue $l(l+d-2)$, $l=0,1,2,\dots$.  Thus, we obtain the radial equation of Klein-Gordon equation in $d$ dimensions by substituting Eq.(\ref{eq2}) into Eq.(\ref{eq1})
\begin{equation}\label{eq3}
-R''(r)-\left({{d-1}\over r}\right)R'(r)+\bigg\{{l(l+D-2)\over r^2}+[M+S(r)]^2-[E-V(r)]^2\bigg\}R({\bf r})=0.
\end{equation}
Writing $R$ as $R(r)=r^{-(D-1)/2}u(r)$ gives the radial equation
\begin{eqnarray}\label{eq4}
-u''(r)&+&\bigg\{{(k-1)(k-3)\over 4r^2}+\big[[M+S(r)]^2-[E-V(r)]^2\big]\bigg\}u(r)=0.
\end{eqnarray}
where $k=d+2l$ and $u(r)$ is the reduced radial wave function satisfying $u(0) = 0.$ 

\section{Brief introduction of the solution method}
\noindent The asymptotic iteration method was introduced \cite{Hakan} to obtain exact and approximate solutions of eigenvalue equations \cite{hall}. The first step in applying this method to solve Schr\"odinger-type equations is to transform these equations, with the aid of appropriate asymptotic forms, to second-order homogeneous linear differential equations of the general form
\begin{equation}\label{eq5}
y^{\prime\prime}=\lambda_0(r)y^\prime+s_0(r)y,
\end{equation}
for which  $\lambda_0(r)$ and $s_0(r)$ are functions in $C^{\infty}$. Here primes denotes the derivatives with respect to  $r$. A key feature of AIM is to note the invariant structure of the right-hand side of (\ref{eq5}) under further differentiation. Indeed, if we differentiate (\ref{eq5}) with respect to $r$, we find that
\begin{equation}\label{eq6}
y^{\prime\prime\prime}=\lambda_1(r)y^\prime+s_1(r)y
\end{equation}
where
$$
\left\{
\begin{array}{l}
\lambda_1=\lambda_0^\prime+s_0+\lambda_0^2\\
s_1=s_0^\prime+s_0\lambda_0.
\end{array}
\right.
$$
Meanwhile the second derivative of (\ref{eq5}) yields
\begin{equation}\label{eq7}
y^{(4)}=\lambda_2(r)y^\prime+s_2(r)y
\end{equation}
for which
$$
\left\{
\begin{array}{l}
\lambda_2=\lambda_1^\prime+s_1+\lambda_0\lambda_1\\
s_2=s_1^\prime+s_0\lambda_1.
\end{array}
\right.
$$
Thus, for the $(n-1)^{th}$ and $(n)^{th}$ derivatives of (\ref{eq5}), $n=1,2,\dots$, we have
\begin{equation}\label{eq8}
\begin{array}{l}
y^{(n+1)}=\lambda_{n-1}(r)y^\prime+s_{n-1}(r)y\\
\\
y^{(n+2)}=\lambda_{n}(r)y^\prime+s_{n}(r)y
\end{array}
\end{equation}
respectively, where
\begin{equation}\label{eq9}
\left\{
\begin{array}{l}
\lambda_{n}= \lambda_{n-1}^\prime+s_{n-1}+\lambda_0\lambda_{n-1}\\
s_{n}=s_{n-1}^\prime+s_0\lambda_{n-1}.
\end{array}
\right.
\end{equation}
In an earlier paper \cite{Hakan} we proved the principal theorem of the asymptotic iteration method (AIM), namely
\vskip0.1true in
\noindent{\bf Theorem 1:} \emph{Given $\lambda_0$ and $s_0$ in $C^\infty$, the differential equation (\ref{eq5}) has the general solution
\begin{eqnarray}\label{eq10}
y(r)&=&\exp\left({-\int\limits^{r}{s_{n-1}\over \lambda_{n-1}} dt}\right) \bigg[C_2+C_1\int\limits^{r}\exp\left(\int\limits^{t}(\lambda_0 + {2s_{n-1}\over \lambda_{n-1}}) d\tau \right)dt\bigg]
\end{eqnarray}
if for some $n>0$
\begin{equation}\label{eq11}
\delta_n=\lambda_n s_{n-1}-\lambda_{n-1}s_n =0.
\end{equation}
}

\noindent Recently, it has been shown \cite{saad} that the termination condition (\ref{eq11}) is necessary and sufficient for the differential equation (\ref{eq5}) to have polynomial-type solutions. In the next section, we shall apply this method to obtain the exact solutions for the relativistic $d$-dimension Klein Gordon equation with Coulomb potentials. The results obtained in the next section are known since the Klein-Gordon equation, in this particular case, reduces into an hypergeometric-type differential equation which has known solutions in terms of confluent hypergeometric function. However, the point is to unify the technique of using AIM to obtain analytic solutions, by first studying the known Coulomb case, and then going on to the more-general Kratzer case.  
\section{The Klein-Gordon equation with Coulomb Potentials in $d$-dimensions} 
\noindent In this section, we use AIM to study the $d$-dimensional Klein-Gordon equation with vector $V(r)$ and scalar $S(r)$ Coulomb potentials \cite{dong}-\cite{ma}, namely
\begin{equation}\label{eq12}
V(r)=-{v\over r},\quad S(r)=-{s\over r}.
\end{equation}
In this case, the Klein-Gordon equation (\ref{eq4}) reads
\begin{equation}\label{eq13}
-u''(r)+\bigg\{{(k-1)(k-3)+4(s^2-v^2)\over 4r^2}-{2(Ms+Ev)\over r}\bigg\}u(r)=(E^2-M^2)u(r).
\end{equation}
In order to solve this Schr\"odinger-like differential equation using AIM, as mentioned earlier, the first step is to transform (\ref{eq13}) into the standard form (\ref{eq5}). We note that the differential equation (\ref{eq13}) has one regular singular point at $r=0$ and an irregular singular point at $r=\infty$. The asymptotic solution of (\ref{eq13}) as $r\rightarrow \infty$ is given by
$u(r)\approx e^{-\sqrt{M^2-E^2}r}$, meanwhile the indicial equation of (\ref{eq13}) at the regular singular point $r=0$ yields 
\begin{equation*}
c^2-c-s^2+v^2-{1\over 4}(k-1)(k-3)=0.
\end{equation*}
Thus the exact solution of (\ref{eq13}) may assume the form
\begin{equation}\label{eq14}
u(r)=r^ce^{br}f(r)
\end{equation}
where $b=-\sqrt{M^2-E^2}$ and 
\begin{equation}\label{eq15}
 c={1\over 2}+ \sqrt{({k\over 2}-1)^2+s^2-v^2},
\end{equation}
because only the negative root yields a regular wave function at $r=0$. Substituting (\ref{eq14}) in (\ref{eq13}), we immediately obtain
\begin{equation}\label{eq16}
f''(r)=-2\left({c\over r}+b\right)f'(r)+
\bigg({-2(Ms+Ev)-2cb\over r}\bigg)f(r).
\end{equation}
This is a confluent hypergeometric differential equation which has a regular singular point at $r=0$ and an irregular singularity at $r=\infty$. The application of AIM is now initiated with $\lambda_0=-2({c\over r}+b)$ and $s_0={-2(Ms+Ev)-2cb\over r}$, and then terminated with the condition (\ref{eq11}).  We find
\begin{equation}\label{eq17}
\delta_n=0\quad\mbox{iff}\quad \prod\limits_{k=0}^n(Ms+Ev+cb+kb)=0
\end{equation}
which in turn yields
\begin{equation}\label{eq18}
Ms+Ev+cb=-nb,\quad n=0,1,2,\dots
\end{equation}
Further, use of (\ref{eq15}) yields the eigenvalue equation
\begin{equation}\label{eq19}
{Ms+Ev\over \sqrt{M^2-E^2}}=n+{1\over 2}+ \sqrt{({k\over 2}-1)^2+s^2-v^2}.
\end{equation}
Upon solving (\ref{eq19}) for $E$, we obtain
\begin{equation}\label{eq20}
E=M\bigg(-{sv\over \beta^2+v^2}\pm {\beta\over \beta^2+v^2} \sqrt{\beta^2+v^2-s^2}\bigg),
\end{equation}
where $\beta={n+{1\over 2}+\sqrt{\big({k\over 2}-1\big)^2+s^2-v^2}}$. The un-normalized wave function can be found using (\ref{eq10}), namely
\begin{equation*}
f_n(r)=\exp\bigg(-\int^r{s_{n-1}(\tau)\over \lambda_{n-1}(\tau)}d\tau\bigg),
\end{equation*}
which, after some algebraic computation, yields
\begin{equation}\label{eq21}
f_n(r)=(-1)^n(c+n)^n(2c)_n {}_1F_1\left(-n;2c;2r\sqrt{M^2-E^2}\right),\quad n=0,1,2,\dots
\end{equation}
up to a multiplicative constant. Here ${}_1F_1$ denotes a confluent hypergeometric function \cite{andrew}
\begin{equation}\label{eq22}
{}_1F_1(a;b;x)=1+{a\over b} x+{a(a+1)\over b(b+1)}{x^2\over 2!}+\dots+{(a)_n\over (b)_n}{x^n\over n!}+\dots=\sum\limits_{k=0}^\infty {(a)_k\over (b)_k}{x^k\over k!},
\end{equation}
and $(a)_k={\Gamma(a+k)/\Gamma(a)}$. Using equations (\ref{eq14}) and (\ref{eq21}), we find that the exact solutions of Klein-Gordon equation with Coulomb potential in arbitrary dimension $d$ is given by
\begin{equation}\label{eq23}
u_n(r)=C_nr^{{1\over 2}+\sqrt{({k\over 2}-1)^2+s^2-v^2}}e^{-\sqrt{M^2-E^2}r} {}_1F_1\left(-n;1+\sqrt{(k-2)^2+4(s^2-v^2)};2r\sqrt{M^2-E^2}\right)
\end{equation}
 The normalization constant $C_n$ can be computed by means of $\int_0^\infty |R(r)|^2r^{D-1}dr=1$, where $R(r)=r^{-(D-1)/2}u(r)$, which, in turn, requires the computation of the definite integral 
\begin{equation}\label{eq24}
I_{nm}(\alpha)=\int\limits_0^\infty \rho^\alpha e^{-\rho} {}_1F_1(-n;\alpha;\rho) {}_1F_1(-m;\alpha;\rho)d\rho
\end{equation}
when $n=m$. Since there is some confusion in the notation used in Refs.(\cite{dong}-\cite{ma}) for the computation of the normalization constant $C_n$ in (\ref{eq22}), we shall present here a detailed computation of the definite integral (\ref{eq24}).
\vskip0.1true in
\noindent{\bf Lemma 1:} For $\alpha>-1$, we have
\begin{eqnarray}\label{eq25}
\int\limits_0^\infty&~&\rho^\alpha e^{-\rho} {}_1F_1(-n;\alpha;\rho){}_1F_1(-m;\alpha;\rho)d\rho=\left\{
\begin{array}{ll}
-{[\Gamma(\alpha)]^2\over \Gamma(\alpha+n-1)}n!&\quad if\quad m=n-1\\
\\
(\alpha+2n){n!\Gamma(\alpha)\over (\alpha)_n} &\quad if\quad m = n\\
\\
-{[\Gamma(\alpha)]^2\over \Gamma(\alpha+n)}(n+1)!&\quad if\quad m=n+1 \\
\\
0&\quad {\rm otherwise}.
\end{array}
\right.
\end{eqnarray}
Proof: Using the series representation of the Confluent hypergeometric function (\ref{eq24}) and observing that $(-n)_k=0$ if $k>n$, we can write (\ref{eq24})
\begin{eqnarray*}
I_{nm}&=&\sum\limits_{i=0}^n\sum\limits_{k=0}^m{(-n)_i(-m)_k\over (\alpha)_i(\alpha)_k ~i!~k!}\int\limits_{0}^\infty \rho^{\alpha+i+k}e^{-\rho}d\rho=\sum\limits_{i=0}^n\sum\limits_{k=0}^m{(-n)_i(-m)_k\over (\alpha)_i(\alpha)_k ~i!~k!}\Gamma(\alpha+i+k+1)\\
&=&\sum\limits_{i=0}^n{(-n)_i\Gamma(\alpha+i+1)\over (\alpha)_i~i!}{}_2F_1(-m,\alpha+i+1;\alpha;1)=\Gamma(\alpha+1)\sum\limits_{i=0}^n{(-n)_i(\alpha+1)_i\over (\alpha)_i~i!}{(-i-1)_m\over (\alpha)_m},\nonumber\\
\end{eqnarray*}
as a consequence \cite{andrew2} of Vandermonde's Theorem  ${}_2F_1(-n,b;c;1)=(c-b)_n/(c)_n$.
The last sum survives for $m=n-1,n,n+1$. Since we are concerned with the case $m=n$, we consider
\begin{eqnarray*}
I_{nn}&=&{\Gamma(\alpha+1)\over (\alpha)_n}\sum\limits_{i=0}^n{(-n)_i(\alpha+1)_i(-i-1)_m\over (\alpha)_i~i!}\\
&=&{\Gamma(\alpha+1)\over (\alpha)_n}\bigg\{{(-n)_{n-1}(\alpha+1)_{n-1}(-n)_n\over (\alpha)_{n-1}(n-1)!}+{(-n)_{n}(\alpha+1)_{n}(-n-1)_n\over (\alpha)_{n}n!}\bigg\}=(\alpha+2n){n!\Gamma(\alpha)\over (\alpha)_n} 
\end{eqnarray*}
as required. The other cases follow similarly.\qed
\vskip0.1true in
\noindent Consequently, we have
\begin{equation}\label{eq26}
\int\limits_0^\infty \rho^\alpha e^{-\rho} \big[{}_1F_1(-n;\alpha;\rho)\big]^2 d\rho=(\alpha+2n){n!\Gamma(\alpha)\over (\alpha)_n}
\end{equation}
The normalization constant in (\ref{eq22}) is then provided by:
$$C_n^{-2}=\bigg({1\over 2\sqrt{M^2-E^2}}\bigg)^{2c+1}(2c+2n){n!\Gamma(2c)\over (2c)_n}.
$$
Consequently, the full normalized wave function (\ref{eq23}) reads
\begin{eqnarray}\label{eq27}
u_n(r)=\sqrt{(2\sqrt{M^2-E^2})^{2c+1}(2c)_n\over 2(c+n)n!\Gamma(2c)}r^ce^{-\sqrt{M^2-E^2}r}{}_1F_1(-n;2c;2r\sqrt{M^2-E^2}),
\end{eqnarray}
where $c={1\over 2}+\sqrt{({k\over 2}-1)^2+s^2-v^2}$. Depending on the values of $s$ and $v$, we may now consider the following three cases:
\begin{itemize}
\item If $v=0$, in this case $s$ and $M$ must have the same sign as a result of (\ref{eq20}), i.e. $s>0$ (since the right-hand side of (\ref{eq19}) is greater than 0). Thus we have
\begin{equation}\label{eq28}
E=\pm M\left(1-{s^2\over \big(n+{1\over 2}+\sqrt{({k\over 2}-1)^2+s^2}\big)^2}\right)^{1/2}.
\end{equation}

\item If $s=0$, then, again using (\ref{eq19}), $E$ and $v$ must have the same sign for $({k\over 2}-1)^2>v^2$. For attractive Coulomb potentials $0<v<{D\over 2}+l-1$, we have the so-called \emph{$\pi$-mesonic atom} \cite{dav}
\begin{equation}\label{eq29}
E=M\left(1+{v^2\over \big(n+{1\over 2}+\sqrt{({k\over 2}-1)^2-v^2}\big)^2}\right)^{-1/2}.
\end{equation}
where $n$ and $l$ are the radial and angular quantum numbers.

\item If $s=v>0$, since $|E|<M$, we have
\begin{equation}\label{eq30}
E=M\left(1-{2v^2\over \big(n+{k\over 2}-{1\over 2}\big)^2+v^2}\right).
\end{equation}
\end{itemize}
\section{The Klein-Gordon equation with Kratzer potentials in $d$-dimensions} 
\noindent In this section, we consider the scalar and vector potentials with the forms
\begin{equation}\label{eq31}
S(r)=-{s_1\over r}+{s_2\over r^2},\quad V(r)=-{v_1\over r}+{v_2\over r^2}.
\end{equation}
The Klein-Gordon equation (\ref{eq4}) then reads
\begin{equation}
-u''+\bigg\{{-2(Ms_1+Ev_1)\over r}+{2Ms_2+s_1^2+2Ev_2-v_1^2+{1\over 4}(k-1)(k-3)\over r^2}+{2(v_1v_2-s_1s_2)\over r^3}+{s_2^2-v_2^2\over r^4}\bigg\}u=(E^2-M^2)u.\\ \label{eq32}
\end{equation}
This differential equation has two irregular singular points at $r=0$ and $r=\infty$. As $r\rightarrow \infty$, the differential equation $u''\approx (M^2-E^2)u$ has a solution given by
$u\approx\exp({-\sqrt{M^2-E^2}~r}).$  As $r\rightarrow 0$, it is clear that the differential equation, after replacing $z={1/r}$ and letting $z\rightarrow \infty$, has a solution
$u\approx\exp({-{\sqrt{s_2^2-v_2^2}~r^{-1}}}).$ Thus we may assume the exact solution of (\ref{eq32}) takes the form
\begin{equation}\label{eq33}
u(r)=e^{-\sqrt{M^2-E^2}r-{\sqrt{s_2^2-v_2^2}\over r}}g(r).
\end{equation}
Substituting this expression in (\ref{eq32}), we obtain
\begin{eqnarray}\label{eq34}
g''(r)&=&-2\big(-{a\over r^2}+b\big)g'(r)+\bigg({-2(Ms_1+Ev_1)\over r}+{2(Ms_2+Ev_2)+s_1^2-v_1^2+{1\over 4}(k-1)(k-3)+2ab\over r^2}\nonumber\\
&+&{2(v_1v_2-s_1s_2)-2a\over r^3}\bigg)g(r),
\end{eqnarray}
where we denote $a=-\sqrt{s_2^2-v_2^2}$ and  $b=-\sqrt{M^2-E^2}.$
\subsection{Equal scalar and vector potentials}
\noindent In order to solve (\ref{eq34}), we consider first the case of equal scalar and vector potentials. i.e. $v_1=s_1=B$ and $v_2=s_2=A$. Then, equation (\ref{eq34}) reduces to 
\begin{eqnarray}\label{eq35}
g''(r)&=&-2bg'(r)+\bigg({-2B(M+E)\over r}+{2A(M+E)+{1\over 4}(k-1)(k-3)\over r^2}\bigg)g(r).
\end{eqnarray}
This equation has a regular singular point at $r=0$ with indicial equation given by
$c^2-c-2A(M+E)-{1\over 4}(k-1)(k-3)=0,$ 
which implies
\begin{equation}\label{eq36}
c={1\over 2}+\sqrt{\big({k\over 2}-1\big)^2+2A(M+E)},
\end{equation}
where we assume $c>0$ (because only the negative root yields a bounded solution at $r=0.$)  With $g(r)=r^cf(r)$, equation (\ref{eq35}) now reads
\begin{eqnarray}\label{eq37}
f''(r)&=&-2\left({c\over r}+b\right)f'(r)+\bigg({-2B(M+E)-2cb\over r}\bigg)f(r).
\end{eqnarray}
By a direct application of AIM with $\lambda_0={c\over r}+b$ and $s_0={-2B(M+E)-2cb\over r}$, the termination condition (\ref{eq11}) implies
$\delta_n=0,~n=0,1,2,3,\dots$ if and only if 
\begin{equation}\label{eq38}
\prod\limits_{k=0}^n \big(B(M+E)+cb+kb\big)=0,\Rightarrow c+n={B(M+E)\over \sqrt{M^2-E^2}}.
\end{equation}
Combining (\ref{eq36}) and (\ref{eq38}), we obtain the eigenvalue equation 
\begin{equation}\label{eq39}
{2B(M+E)\over \sqrt{M^2-E^2}}=2n+1+\sqrt{\big({k}-2\big)^2+8A(M+E)},
\end{equation}
which generalizes the results of \cite{qiang} for the $3$-dimension case. Although we can solve Eq.(\ref{eq39}) exactly for $E$, the expression for $E$ is rather complicated. 
We therefore write Eq.(\ref{eq39}) as
\begin{equation}\label{eq40}
2B\sqrt{2M-\beta^2}-\beta\bigg(2n+1+\sqrt{\big({k}-2\big)^2+8A(2M-\beta^2)}\bigg)=0
\end{equation}
and expand it as a power series in $\beta=\sqrt{M-E}$.  Thus we have
\begin{equation}\label{eq41}
2B\sqrt{2M}+(-2n-1-\sqrt{(k-2)^2+16AM})\beta-{B\over\sqrt{2M}}\beta^2+\dots=0.
\end{equation}
As an approximation, if we take only the term in $\beta,$  and solve the energy equation, we obtain the non-relativistic energy \cite{landau}
\begin{equation}\label{eq42}
E'=E-M=-\beta^2=-{8MB^2\over (2n+1+\sqrt{(k-2)^2+16AM})^2}.
\end{equation}
The exact solutions $f_n(r)$ of (\ref{eq37}) can be obtained using (\ref{eq10}) to yield
\begin{eqnarray}\label{43}
f_n(r)&=&(-1)^n(c+n)^n(2c)_n{}_1F_1(-n;2c;2\sqrt{M^2-E^2}r)
\end{eqnarray}
up to a constant. Consequently, the exact solutions (\ref{eq33}) of (\ref{eq32}), in the case equal scalar and vector potentials, are given by 
\begin{equation}\label{44}
u_n(r)=N_n r^c e^{-\sqrt{M^2-E^2}r}{}_1F_1(-n;2c;2r\sqrt{M^2-E^2}),\quad\quad n=0,1,2,\dots
\end{equation}
where the normalization constant (implied by Lemma~1) now becomes
\begin{equation}\label{45}
N_n=\sqrt{(2\sqrt{M^2-E^2})^{2c+1}(2c)_n\over 2(c+n) n!\Gamma(2c)},
\end{equation}
where $c$ is given by (\ref{eq36}). 
We note that in the case $A=0$, equation (\ref{eq39}) yields
\begin{equation}\label{46}
E=M\bigg(1-{2B^2\over \big(n+{k\over 2}-{1\over 2}\big)^2+B^2}\bigg),
\end{equation}
which is in complete agreement with our results for the Coulomb potential.
\subsection{Unequal scalar and vector potentials}
\noindent In the case of different scalar and vector potentials, we have $a\neq 0$ and therefore we have to approach equation (\ref{eq34}) directly. The equation, however, has two irregular singular points, at $r=0$ and $r=\infty$, and  the standard techniques of solving differential equations cannot be applied. AIM, however, has the advantage that it can be applied directly to obtain exact solutions under certain conditions on the potential parameters of (\ref{eq34}). Indeed, with
$
\lambda_0(r)= -2\big(-{a\over r^2}+b)$ and $s_0(r)={-2(Ms_1+Ev_1)\over r}+{2(Ms_2+Ev_2)+s_1^2-v_1^2+{1\over 4}(k-1)(k-3)+2ab\over r^2}+{2(v_1v_2-s_1s_2)-2a\over r^3},
$
the termination relation (\ref{eq11}) yields, for $\delta_n=0$, where $n=1,2,\dots$ is the \emph{iteration} number, the following conditions
\begin{equation}\label{eq47}
\left\{
\begin{array}{l}
2nb=-2(Ms_1+Ev_1),\\
\\
n(n-1)=2(Ms_2+Ev_2)+s_1^2-v_1^2+{1\over 4}(k-1)(k-3)+2ab\\
\\
-2na=2(v_1v_2-s_1s_2)-2a.
\end{array}
\right.
\end{equation}
Furthermore, the exact solutions under the constraints (\ref{eq47}), using the AIM expression (\ref{eq10}), are given by
\begin{equation}\label{eq48}
g_n(r)=r^n\Rightarrow 
u(r)=C_nr^{n}e^{-\sqrt{M^2-E^2}r-{\sqrt{s_2^2-v_2^2}\over r}}, \quad n=1,2,\dots.
\end{equation}
The normalization constants $C_n$ of (\ref{eq48}) can be computed by the standard identity \cite{grad} (in which $K_{\nu}$ is the modified Bessel function of second kind of order $\nu$) 
\begin{equation}\label{eq49}
\int\limits_0^\infty r^{\mathcal C}\exp[-{\mathcal B}r-{{\mathcal A}\over r}]dr=2\bigg({{\mathcal A}\over {\mathcal B}}\bigg)^{{\mathcal C}+1\over 2}K_{-{\mathcal C}-1}(2\sqrt{{\mathcal A}{\mathcal B}}),\quad\quad {\mathcal A}>0, {\mathcal B}>0.
\end{equation}
which implies that 
\begin{equation}\label{eq50}
C_n^{-2}=2\left({s_2^2-v_2^2\over M^2-E^2}\right)^{{1+n\over 4}}K_{-n-1}(2(s_2^2-v_2^2)^{1\over 4}(M^2-E^2)^{1\over 4}).
\end{equation}
That is to say, the exact solutions for the differential equation (\ref{eq34}) under the constraints (\ref{eq47}) are given by
\begin{equation}\label{eq51}
u_n(r)=\left({2\left({s_2^2-v_2^2\over M^2-E^2}\right)^{{1+n\over 4}}K_{-n-1}(2(s_2^2-v_2^2)^{1\over 4}(M^2-E^2)^{1\over 4})}\right)^{-{1\over 2}}
r^{n}e^{-\sqrt{M^2-E^2}r-{\sqrt{s_2^2-v_2^2}\over r}}.
\end{equation}
Further, we note that the first and the third equations of (\ref{eq47}) yield
\begin{equation}\label{eq52}
\left\{
\begin{array}{l}
n={Ms_1+Ev_1\over \sqrt{M^2-E^2}}\\
\\
n={v_1v_2-s_1s_2\over \sqrt{s_2^2-v_2^2}}+1,
\end{array}
\right.
\end{equation}
This, in turn, inplies
\begin{equation}\label{eq53}
\sqrt{s_2^2-v_2^2}\sqrt{M^2-E^2}=(Ms_1+Ev_1)\sqrt{s_2^2-v_2^2}-(v_1v_2-s_1s_2)\sqrt{M^2-E^2}.
\end{equation}
Thus, by mean of the second equation of (\ref{eq47}), the eigenvalue equation is
\begin{eqnarray}
{Ms_1+Ev_1\over \sqrt{M^2-E^2}}={1\over 2}+\sqrt{({k\over 2}-1)^2+2(Ms_1+Ev_1)\sqrt{s_2^2-v_2^2}-2(v_1v_2-s_1s_2)\sqrt{M^2-E^2}
+2(Ms_2+Ev_2)+s_1^2-v_1^2}.\nonumber\\
\label{eq54}
\end{eqnarray}
Some important remarks are in order:
\begin{itemize}
\item The exact solutions $u_n(r)$, $n=1,2,\dots$ are nodeless for all $n$. Here, $n$ represents the AIM iteration number. Thus, for all the values of the potential parameters that satisfy (\ref{eq51})-(\ref{eq52}), the corresponding solution represents a ground-state wave function. Note that each operator now acts on a different parameter space. 
\item We notice for $n=1$, the second equation of (\ref{eq51}) implies $v_1v_2=s_1s_2$ and since $s_2>v_2$, we must have $v_1>s_1$ which in turn implies $-s_1>-v_1$. Thus, the requirement for bound states $S(r)>V(r)$ is satisfied. It is also clear that, this is the case for all $n\ge 1$.
\item If $s_2=v_2=0$, we obtain 
$u_0(r)=C_0r^{{1\over 2}+\sqrt{({k\over 2}-1)^2+s_1^2-v_1^2}}e^{-\sqrt{M^2-E^2}r}$
and
\begin{eqnarray*}
{Ms_1+Ev_1\over \sqrt{M^2-E^2}}&=&{1\over 2}+\sqrt{({k\over 2}-1)^2+s_1^2-v_1^2},
\end{eqnarray*}
which reduces to the ground state solution described for Coulomb potential as mentioned earlier.
\item In the case of a pure scalar potential, i.e. $V(r)=0$, we have $v_1=v_2=0$, $S(r)=-{s_1\over r}+{s_2\over r^2}>0$,
\begin{eqnarray*}
u_0(r)&=&C_0r^{{1\over 2}+\sqrt{({k\over 2}-1)^2+2Ms_1s_2+2s_1s_2\sqrt{M^2-E^2}+2Ms_2+s_1^2}} e^{-\sqrt{M^2-E^2}r-{s_2\over r}},
\end{eqnarray*}
and
\begin{eqnarray*}
{Ms_1\over \sqrt{M^2-E^2}}&=&{1\over 2}+\bigg(({k\over 2}-1)^2+2Ms_1s_2+s_1s_2\sqrt{M^2-E^2}
+2Ms_2+s_1^2\bigg)^{1/2}.
\end{eqnarray*}
If we further assume that $s_2=0$, we obtain
\begin{eqnarray*}
{Ms_1\over \sqrt{M^2-E^2}}&=&{1\over 2}+\bigg(({k\over 2}-1)^2+s_1^2\bigg)^{1/2},
\end{eqnarray*}
which again agrees with the result for the Coulomb potential for this particular case.
\end{itemize}

\noindent In order to move beyond the ground state solutions, equation (\ref{eq51}) suggests that the exact solutions $u(r)$ of (\ref{eq34}) take the form
\begin{equation}\label{eq55}
u(r)=r^ce^{-\sqrt{M^2-E^2}r-{\sqrt{s_2^2-v_2^2}\over r}}f(r).
\end{equation}
Substituting this expression in (\ref{eq33}), we obtain
\begin{eqnarray}\label{eq56}
f''(r)&=&-2\big({c\over r}-{a\over r^2}+b\big)f'(r)+\bigg({-2(Ms_1+Ev_1)-2cb\over r}+{2(Ms_2+Ev_2)+s_1^2-v_1^2+{1\over 4}(k-1)(k-3)-c^2+c+2ab\over r^2}\nonumber\\
&+&{2(v_1v_2-s_1s_2)-2a+2ca\over r^3}\bigg)f(r).
\end{eqnarray}
where again $a=-\sqrt{s_2^2-v_2^2}$ and  $b=-\sqrt{M^2-E^2}.$ We may now choose $c$ such that
\begin{equation}\label{eq57}
v_1v_2-s_1s_2-a+ca=0\Rightarrow c={v_1v_2-s_1s_2\over \sqrt{s_2^2-v_2^2}}+1.
\end{equation}
which in turn reduces Eq.(\ref{eq56}) to
\begin{eqnarray}\label{eq58}
f''(r)&=&-2\left({c\over r}-{a\over r^2}+b\right)f'(r)+
\bigg({-2(Ms_1+Ev_1)-2cb\over r}+{G\over r^2}\bigg)f(r),
\end{eqnarray}
where we denote
\begin{eqnarray}\label{eq59}
G=2(Ms_2+Ev_2)+s_1^2-v_1^2+{1\over 4}(k-1)(k-3)-c^2+c+2ab.
\end{eqnarray}
Although (\ref{eq58}) still has an irregular singular point at $r=0$, the direct application of AIM with $\lambda_0=-2\big({c\over r}-{a\over r^2}+b)$ and $s_0={-2(Ms_1+Ev_1)-2cb\over r}+{G\over r^2}$, implies, by means of the termination condition (\ref{eq11}), that
\begin{equation}\label{eq60}
Ms_1+Ev_1+cb=-nb\quad\Rightarrow\quad c+n={Ms_1+Ev_1\over \sqrt{M^2-E^2}},\quad\quad n=0,1,2,\dots,
\end{equation}
where for $n=0,1,2,\dots$ the following constraints on $G$ must hold
\begin{eqnarray*}
\mathcal G_0&:=&G=0\nonumber\\
\mathcal G_1&:=&G^2-2cG-4ab=0\nonumber\\
\mathcal G_2&:=&G^3-2(3c+1)G^2+4(2c^2+c-4ba)G+16ab(2c+1)=0\\
\mathcal G_3&:=&G^4-4(3c+2)G^3+4(11c^2+3+13c-10ab)G^2+(192ab-72c^2+240cab-24c-48c^3)G\\
&-&144ba+144a^2b^2-432cab-288c^2ab=0\\
\dots
\end{eqnarray*}
and so on, for higher iteration numbers.
\subsection{The case $\mathcal G_0=0$} 
\noindent In the case $G=0$, we have
\begin{eqnarray}\label{eq61}
c={1\over 2}+\sqrt{({k\over 2}-1)^2+2\sqrt{s_2^2-v_2^2}\sqrt{M^2-E^2}+2(Ms_2+Ev_2)+s_1^2-v_1^2}
\end{eqnarray}
as a root of  (\ref{eq59}) and, again, $c>0$ because only the negative root yields a regular wave function at $r=0.$ 
On the other hand, equation (\ref{eq60}) implies
\begin{equation}\label{eq62}
c={Ms_1+Ev_1\over \sqrt{M^2-E^2}},
\end{equation}
where now (\ref{eq57}) gives
\begin{equation}\label{eq63}
\sqrt{s_2^2-v_2^2}\sqrt{M^2-E^2}=(Ms_1+Ev_1)\sqrt{s_2^2-v_2^2}-(v_1v_2-s_1s_2)\sqrt{M^2-E^2}.
\end{equation}
Therefore, the ground-state energy equation in $d$-dimensions is given by 
\begin{eqnarray}
{Ms_1+Ev_1\over \sqrt{M^2-E^2}}={1\over 2}+\sqrt{({k\over 2}-1)^2+2(Ms_1+Ev_1)\sqrt{s_2^2-v_2^2}-2(v_1v_2-s_1s_2)\sqrt{M^2-E^2}
+2(Ms_2+Ev_2)+s_1^2-v_1^2},\nonumber\\
\label{eq64}
\end{eqnarray}
which in complete agreement with equation (\ref{eq54}).

\subsection{The case $\mathcal G_1=0$} 
\noindent In this case, the constraint $\mathcal G_1:=G^2-2cG-4ab=0$ implies $G=c\pm \sqrt{c^2+4ab},$
which implies by (\ref{eq59}) for $G>0$, that
\begin{equation}\label{eq65}
\sqrt{c^2+4ab}=2(Ms_2+Ev_2)+s_1^2-v_1^2+{1\over 4}(k-1)(k-3)-c^2+2ab.
\end{equation}
In order to simplify the notation, we denote 
\begin{equation}\label{eq66}
\mu=2(Ms_2+Ev_2)+s_1^2-v_1^2+{1\over 4}(k-1)(k-3)+2ab
\end{equation}
and solve (\ref{eq65}) for $c>0$ to obtain
\begin{equation}\label{eq67}
c={1\over 2}\sqrt{2+4\mu+2\sqrt{4\mu+1+16ab}}.
\end{equation}
Meanwhile, (\ref{eq57}) and (\ref{eq60}) imply, for $ab=\sqrt{M^2-E^2}\sqrt{s_2^2-v_2^2}$~, that 
\begin{equation}\label{eq68}
ab={1\over 2}(Ms_1+Ev_1)\sqrt{s_2^2-v_2^2}-{1\over 2}(v_1v_2-s_1s_2)\sqrt{M^2-E^2}.
\end{equation}
The eigenenergy is then given, from (\ref{eq67}), by the equation
\begin{equation}\label{eq69}
{Ms_1+Ev_1\over \sqrt{M^2-E^2}}={1\over 2}\sqrt{2+4\mu+2\sqrt{4\mu+1+16ab}}-1,
\end{equation}
where $\mu$ and $ab$ are given by (\ref{eq66}) and (\ref{eq68}), respectively. Further, the exact solution of (\ref{eq56}) and $G=c+ \sqrt{c^2+4ab}$ becomes 
\begin{equation}\label{eq70}
f_1(r)=r-{2a\over G}
\end{equation}
up to a constant. Thus
\begin{equation}\label{eq71}
u_1(r)=C_1\big(r-{2a\over G}\big)r^ce^{-\sqrt{M^2-E^2}r-{\sqrt{s_2^2-v_2^2}\over r}},
\end{equation}
where the normalization constant $C_1$ can be computed by means of (\ref{eq49}).
\subsection{The case $\mathcal G_2=0$} 
\noindent In this case, we have for
$Ms_1+Ev_1+cb=-2b
$ that
$\mathcal G_2:=G^3-2(3c+1)G^2+4(2c^2+c-4ba)G+16ab(2c+1)=0$,
where $G$ is given by (\ref{eq59}). The eigenvalue equation is then given by the root of this equation along with $G$ given by (\ref{eq59}) and 
\begin{equation}\label{eq72}
ab={1\over 3}(Ms_1+Ev_1)\sqrt{s_2^2-v_2^2}-{1\over 3}(v_1v_2-s_1s_2)\sqrt{M^2-E^2},
\end{equation}
in a similar fashion to the previous case. After some algebraic computations, we obtain for the exact solution that
\begin{equation}\label{eq73}
f_2(r)=(r-{G-4c-2\over 4b}+{1\over 4b}\sqrt{16ab+8c+4+4cG-G^2})(r-{G-4c-2\over 4b}-{1\over 4b}\sqrt{16ab+8c+4+4cG-G^2}).
\end{equation}
\subsection{The general case $\mathcal G_n:=0$} 
\noindent The structure of the wave functions for the cases $\mathcal G_0=0$, $\mathcal G_1=0$, and $\mathcal G_2=0$, etc., allows us to formulate a possible general solution of the Klein-Gordon equation in the case of unequal scalar and vector Kratzer potentials for arbitrary $n$ and $d$. For this purpose, we assume that the general form of the exact solution of the differential equation (\ref{eq59}) takes the form of a monic polynomial
\begin{equation}\label{eq74}
f_n(r)=\prod\limits_{k=1}^n (r-\sigma_k)=\sum\limits_{k=0}^n a_kr^k
\end{equation}
whose coefficients $\{a_k\}_{k=0}^n$ are the elementary symmetric polynomials \cite{macdonald}
\begin{equation}\label{eq 79}
\left\{ \begin{array}{l}
 a_n~~~=1, \\ \\
a_{n-1}=-\sum\limits_{1\leq k\leq n}\sigma_k,\\ \\
a_{n-2}=\sum\limits_{1\leq i<j\leq n}\sigma_i\sigma_j,\\ \\
a_{n-3}=-\sum\limits_{1\leq i<j<k\leq n}\sigma_i\sigma_j\sigma_k,\\ 
\dots~~=\dots \\ 
a_k~~~=\sum\limits_{1\leq j_1<j_2<\dots<j_k\leq n}\sigma_{j_1}\dots \sigma_{j_k}\\ 
\dots~~=\dots \\ 
a_0~~~=(-1)^n\prod\limits_{k=1}^n \sigma_k
\end{array} \right.
\end{equation}
On substituting (\ref{eq74}) into the differential equation (\ref{eq58}), we find that the coefficients $\{a_k\}_{k=0}^n$ must satisfy the following set of linear equations:
\begin{equation}\label{eq76}
\left\{ \begin{array}{ll}
Ga_0+2aa_1=0,&~ \\ \\
4aa_2+(G-2c)a_1+2nba_0=0,&~ \\ \\
\left[n(n-1)+2cn-G\right]~a_n=2ba_{n-1},&~\\ \\ 
2a(k+3)a_{k+3}+(G-2c(k+2)-(k+1)(k+2))a_{k+2}+(F-2b(k+1))a_{k+1}=0&, \mbox{ for\quad $0\leq k\leq n-3$.}
\end{array} \right.
\end{equation}
In order to understand the application of these formulae, we consider the case of $n=3$: equation (76) implies
\begin{equation}\label{eq77}
\left\{ \begin{array}{ll}
a_2=-{1\over 4a}(G-2c-{12ab\over G})a_1,&~ \\ \\
a_3={2b\over 6+6c-G}a_2,&~\\ \\ 
6aa_3+(G-4c-2)a_2+4ba_1=0&, \mbox{ since \quad $k=0$.}
\end{array} \right.
\end{equation}
The solution of this system, for $a_1\neq 0$, gives the condition  $\mathcal G_3$ on $G$ that reads
\begin{eqnarray*}\label{eq82}
G^4&-&4(3c+2)G^3+4(11c^2+3+13c-10ab)G^2+(192ab-72c^2+240cab-24c-48c^3)G\\&-&144ba+144a^2b^2-432cab-288c^2ab=0
\end{eqnarray*}
That is to say, the same condition that we obtain if we apply AIM to (\ref{eq58}) with three iterations, i.e $n=3$. Equations (\ref{eq76}) are the full set of restrictions on the parameters of the Kratzer potentials (\ref{eq31}) in the case of unequal scalar and vector parts.
\section{Conclusion}
\noindent In this paper, the Klein Gordon equation in arbitrary dimension has been solved exactly for the bound states corresponding to Coulomb or Kratzer scalar $S(r)$ and vector $V(r)$ potentials. When both potentials are Coulombic, equal or not, we find all the analytic solutions. When both potentials are of Kratzer type, we find the exact solutions when $S(r) = V(r);$ when the potentials are unequal and the scalar potential dominates, we generate exact solutions under certain specific conditions on the potential parameters. Furthermore, a general solution is found in terms of a monic polynomial whose coefficients form a set of elementary symmetric polynomials. Our method of solution is based on the recently-introduced  Asymptotic Iteration Method.  This approach has the advantage of simplicity in the exact cases, and flexibility, leading to approximations, when exact solutions are not attainable.
\section*{Acknowledgments}
\noindent Partial financial support of this work under Grant Nos. GP3438 and GP249507 from the 
Natural Sciences and Engineering Research Council of Canada is gratefully 
acknowledged by two of us (respectively [RLH] and [NS]).

\end{document}